\documentclass[aps,pra,10pt,notitlepage,twocolumn]{revtex4-1}
\usepackage[dvips]{graphicx}
\usepackage{amssymb,amsfonts}
\usepackage{amsmath}
\usepackage{dsfont}

\newcommand{\tr}{\mathrm{Tr}}
\newcommand{\bra}[1]    {\langle #1|}
\newcommand{\ket}[1]    {\left| #1 \right\rangle}
\newcommand{\braket}[2]{\langle #1 | #2 \rangle }
\newcommand{\re}{\mathrm{Re}}
\newcommand{\im}{\mathrm{Im}}
\newcommand{\x}{{\mathbf r}}

\begin{document}

\author{T. Wasak$^1$, J. Chwede\'nczuk$^1$, L. Pezz\'e$^2$ and A. Smerzi$^2$}
\affiliation{
  $^1$Faculty of Physics, University of Warsaw, ul.\ Ho\.{z}a 69, PL--00--681 Warszawa, Poland\\ 
  $^2$QSTAR, INO-CNR and LENS,  Largo Enrico Fermi 2, 50125 Firenze, Italy 
}

\title{Optimal measurements in phase estimation: simple examples}

\begin{abstract}
  We identify optimal measurement strategies for phase estimation in different scenarios.
  For pure states of a single qubit, we show that optimal measurements form a broad set parametrized with a continuous variable.
  When the state is mixed this set reduces to merely two possible measurements. 
  For two qubits, we focus on the symmetric Werner state. We find an optimal measurement and show that estimation
  from the population imbalance is optimal only if the state is pure. Finally, for a pure state of $N$ qubits, 
  we finds under which conditions the estimation from the 
  full $N$-body correlation and from the population imbalance are optimal. 
\end{abstract}
\maketitle

\section{Introduction and Outline}

Quantum interferometry aims at estimating an unknown phase $\theta$ 
with the smallest possible uncertainty $\Delta \theta$ \cite{Paris,GiovannettiNATPHOT2011}.
The estimation scheme generally consists of three steps: a quantum state $\hat\rho_0$ of $N$ particles is prepared,
phase shifted by the interferometric device and finally measured.
The procedure is repeated with $m$ copies of $\hat\rho_0$.
The measurement results are used to infer $\theta$ as the value of the estimator $\theta_{\rm est}$.
According to the Cramer-Rao Lower Bound (CRLB), the mean square fluctuation of the estimator, 
giving the sensitivity of the phase estimation, 
is bounded by the inverse of the Fisher Information, $\Delta\theta\geqslant\frac1{\sqrt{m F}}$ \cite{helstrom,cramer}. 
The value of $F$ is determined by all the three steps of the interferometric sequence, namely the state $\hat\rho_0$, the phase acquisition and the estimation. 
When the $N$ particles forming $\hat\rho_0$ are uncorrelated, then the Fisher Information is bounded by the Shot-Noise Limit (SNL), 
i.e. $F\leqslant N$ \cite{pezze, GiovannettiPRL2006}.
Nevertheless the SNL can be surpassed when $\hat\rho_0$ is entangled \cite{sorensen,pezze,HyllusPRA2012, TothPRA2012}, 
up to the Heisenberg limit $F=N^2$ when the phase is imprinted on the
NOON state $\ket\psi=\frac1{\sqrt 2}(\ket{N0}+\ket{0N})$. 
Sensitivities below the SNL has been experimentally achieved with ions \cite{ions},
photons \cite{photons}, cold \cite{atoms} and ultracold \cite{esteve, riedel, gross, app, berrada, smerzi} atoms.
However, the choice of potentially useful entangled input state must be accompanied with a proper measurement, such that gives highest
possible value of $F$ and thus -- via the CRLB -- minimal $\Delta\theta$. 
The maximal value of $F$ (with respect to all possible measurements) for a given $\hat\rho_0$ is called the 
Quantum Fisher Information (QFI) \cite{helstrom,braun} and will be denoted by $F_Q$. 
Although in many cases, especially when $\hat\rho_0$ is a pure state, it is relatively simple to calculate the $F_Q$, it is usually difficult to tell which is the optimal measurement
such that gives $F=F_Q$. 

In this work we identify optimal measurements in different two-mode interferometric systems. We start in Sec.~\ref{single} with the simplest possible two-mode object, which is a single
qubit. We find optimal measurements, when the interferometric transformation is a rotation of the state on the Bloch sphere by an unknown angle $\theta$. While for pure states 
$\hat\rho_0$, there is a whole continuity of optimal estimation strategies, they boil down to only two possibilities when $\hat\rho_0$ is mixed. For two qubits such general 
considerations are not possible anymore, so in Sec.~\ref{wern} we use a particular example, namely the symmetric Werner state. We find the optimal measurement and discuss the precision
which can be reached when the phase is estimated from the population imbalance between the two modes. Finally, in Sec.~\ref{Npure} we show under which conditions the estimation from the population
imbalance or the $N$-th order correlation function is optimal, using $N$-qubit pure state at the input. %% Finally, in Sec.~\ref{Nmix} we introduce some phase noise into the evolution of the
We conclude in Sec.~\ref{concl}. 

\section{Single qubit}
\label{single}

We begin our analysis of the optimal estimation strategies with a most basic two-mode object, namely a single qubit, which is rotated on the Bloch sphere
by an unknown angle $\theta$. As will be shown below, in this case a full family of optimal measurements can be identified, both for pure and mixed states.

\subsection{General formulation}

The density matrix of a single qubit can be represented as a combination of a scalar and three Pauli matrices
\begin{equation}
  \hat\rho_{\rm in}=\frac12(\hat{\mathds1}+\vec s_{\rm in}\cdot\vec{\hat\sigma})
\end{equation}
Depending on the length $s_{\rm in}$ of the vector $\vec s_{\rm in}$, the state is either pure ($s_{\rm in}=1$) or mixed ($s_{\rm in}<1$).

In the scenario considered here, this generic initial state $\hat\rho_{\rm in}$ undergoes a unitary phase-dependent interferometric transformation $\hat U(\theta)$. 
To establish the analogy between the $N$-qubit Mach-Zehnder Interferometer (MZI) and a single-qubit operation, we notice that the former case
is represented by a unitary transformation 
\begin{equation}\label{trans_mzi}
  \hat U_{\rm mzi}(\theta)=e^{-i\theta\hat J_y}. 
\end{equation}
Here $\hat J_y=\sum_{i=1}^N\frac{\hat\sigma_y^{(i)}}2$
is a sum of $y$-component Pauli matrices acting on the $i$-th particle. Clearly, a single-qubit analogy of the MZI is
\begin{equation}\label{1qubit_transform}
  \hat U(\theta)=\exp\left[-i\theta\frac{\hat\sigma_y}2\right],
\end{equation}
which transforms the initial density matrix into
\begin{equation}\label{rotated}
  \hat\rho=\hat U(\theta)\,\hat\rho_{\rm in}\,\hat U^\dagger(\theta)=\frac12(\hat{\mathds1}+\vec{s}\cdot\vec{\hat\sigma}),
\end{equation}
where the three components of the rotated vector are
\begin{subequations}\label{rotated}
  \begin{eqnarray}
    &&s_x=s_{{\rm in},x}\cos\theta+s_{{\rm in},z}\sin\theta\\
    &&s_y=s_{{\rm in},y}\\
    &&s_z=s_{{\rm in},z}\cos\theta-s_{{\rm in},x}\sin\theta
  \end{eqnarray}
\end{subequations}
Note that we have omitted the explicit dependence of $\hat\rho$ on the phase $\theta$ -- in order to simplify the notation.

\subsection{Classical and quantum Fisher information}

We find the optimal estimation phase estimation strategies by considering the broadest set of measurements allowed by quantum mechanics. The elements of this set
are described in terms of the POVMs (Positive-Operator Valued Measures) \cite{helstrom}, 
which are self-adjoint and have non-negative eigenvalues. For a single qubit, any POVM can be represented by the following operator
\begin{equation}\label{povm}
  \hat E_{\vec q}=\gamma_{\vec q}\,(\hat{\mathds1}+\vec q\cdot\vec{\hat\sigma}),
\end{equation}
where $|\vec q| \leqslant 1$ and furthermore
\begin{equation}
  \int\! d\vec q\,\gamma_{\vec q}=1\ \ \ \mathrm{and}\ \ \   \int\! d\vec q\,\gamma_{\vec q}\,\vec q=0.
\end{equation}
The trace of the product of the POVM (\ref{povm}) and the density matrix gives the probability for finding the qubit aligned along $\vec q$ on the Bloch sphere, 
\begin{equation}
  p(\vec q\,|\theta)=\tr\left[\hat\rho\,\hat E_{\vec q}\right].
\end{equation}
The precision of the phase estimation from a series of $m$ measurements is limited by the Cramer-Rao Lower Bound \cite{helstrom,cramer}
\begin{equation}\label{crlb}
  \Delta\theta\geqslant\frac1{\sqrt m}\frac1{\sqrt F},
\end{equation}
where $F$ is called the Classical Fisher Information (CFI) and is equal to
\begin{equation}\label{cfi}
  F=\int\! d\vec q\,\frac1{p(\vec q\,|\theta)}\left(\frac{\partial\, p(\vec q\,|\theta)}{\partial\theta}\right)^2.
\end{equation}
The CFI, which is a measure of information about $\theta$ contained in $\hat\rho$, depends on the particular choice of measurement --
some methods of estimating $\theta$ are better than other. The {\it optimal} measurements are those which give the maximal value of $F$, 
called the Quantum Fisher Information (QFI) \cite{braun} -- and through Eq.~(\ref{crlb}) maximal precision of phase estimation. 

The maximization procedure of (\ref{cfi}) can be performed analytically \cite{braun} and the resulting QFI for  unitary transformations, denoted as $F_Q$ is
\begin{equation}\label{fq_gen}
  F_Q=2\sum_{j,k}\frac{(p_j-p_k)^2}{p_j+p_k}\big|\langle j|\hat h|k\rangle\big|^2.
\end{equation}
Here $\hat h$ is a generator of the phase transformation, which according to Eq.~(\ref{1qubit_transform}) is $\hat h=\frac{\hat\sigma_y}2$. The ket $|k\rangle$ denotes the $k$-th eigen-vector of the density
matrix $\hat\rho$ from Eq.~(\ref{rotated}) with a corresponding eigen-value $p_k$. 
For this 2$\times$2 matrix, the eigen-problem is easily solved and we obtain
\begin{equation}\label{qfi}
  F_Q=s_x^2+s_z^2.
\end{equation}
Using the expressions from  Eq.~(\ref{rotated}), we note that the QFI does not depend on $\theta$ and the maximum $F_Q=1$ is achieved with pure states with $s_y=0$. 

\subsection{Optimal measurements}

Next, we determine which are the optimal measurements, which give (\ref{qfi}). As shown in \cite{braun} the POVM is optimal if and only if it satisfies the condition
\begin{equation}\label{opt_cond}
  \hat E_{\vec q}\,\hat\rho=\lambda_{\vec q}\,\hat E_{\vec q}\,\hat{\mathcal{L}}_{\hat\rho}\,\hat\rho
\end{equation}
with $\lambda_{\vec q}\in\mathbb{R}$. The super-operator $\hat{\mathcal{L}}_{\hat\rho}$ is called the symmetric logarithmic derivative and is defined by a relation
\begin{equation}\label{defL}
  \frac12\left(\hat\rho\hat{\mathcal{L}}_{\hat\rho}+\hat{\mathcal{L}}_{\hat\rho}\hat\rho\right)=\partial_\theta\hat\rho.
\end{equation}
The optimal measurements are found in two steps. First, by declaring the general form of the symmetric logarithmic derivative 
\begin{equation}
  \hat{\mathcal{L}}_{\hat\rho}=\alpha\hat{\mathds1}+\vec s_\perp\vec{\hat\sigma}
\end{equation}
and using the definition (\ref{defL}) we get that $\alpha=0$ and $\vec s_\perp=s_z\vec e_x-s_x\vec e_z$. This result justifies the notation, since $\vec{s}\cdot\vec{s}_\perp=0$. 

\begin{figure}[htb!]
  \includegraphics[clip, scale=0.42]{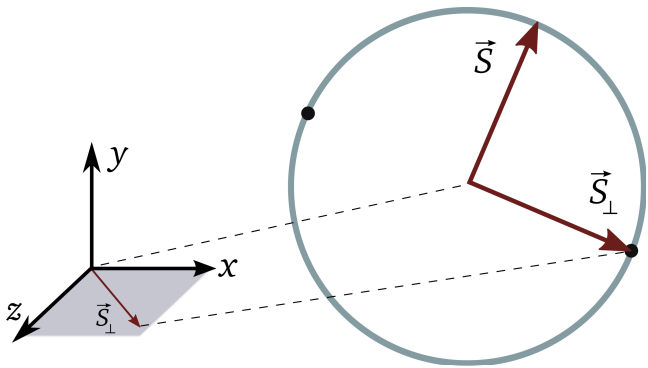}
  \caption{(Color online) Schematic representation of the directions of vectors $\vec q$ for which the POVMs (\ref{povm}) are optimal. The vector $\vec s_\perp$ is orthogonal to $\vec s$ and lies
    in the $x-z$ plane. For pure states, when $|\vec s|=1$, the optimal POVMs lie on a unit circle (blue) in the plane spanned by $\vec s$ and $\vec s_\perp$. For mixed states, for which $|\vec s|<1$,
    only two vectors on circle, denoted by black dots, form the set of optimal POVM. 
  }\label{vecs_fig}
\end{figure}

In the second step we insert $\hat{\mathcal{L}}_{\hat\rho}$ into (\ref{opt_cond}), and use the general parametrization of the POVM (\ref{povm}). 
By comparing the scalar and vector parts and then the real and imaginary parts we obtain a set of equations
\begin{subequations}\label{conds}
  \begin{eqnarray}
    &&\vec q\cdot\left(\vec{s}_\perp\times\vec{s}\right)=0\label{cond1}\\
    &&\lambda_{\vec q}=\frac{1+\vec q\cdot\vec{s}}{\vec q\cdot\vec{s}_\perp}\label{cond2}\\
    &&\vec q+\vec{s}=\lambda_{\vec q}\left[\vec{s}_\perp-\vec q\times\left(\vec{s}_\perp\times\vec{s}\right)\right]\label{cond3}\\
    &&\vec q\times\vec{s}=\lambda_{\vec q}\left[\vec{s}_\perp\times\vec{s}+\vec q\times\vec{s}_\perp\right].\label{cond4}
  \end{eqnarray}
\end{subequations}
From (\ref{cond1}) we deduce that $\vec q$ lies in the plane spanned by vectors $\vec{s}$ and $\vec{s}_\perp$, so it can be written as 
$\vec q=q_1\vec{e}_s+q_2\vec{e}_{s_\perp}$. Here, $\vec{e}_s$ and $\vec{e}_{s_\perp}$ are unit vectors pointing into directions $\vec{s}$ and $\vec{s}_{\perp}$. 
This observation reduces the set of eight equations (\ref{conds}) to 
\begin{subequations}\label{set}
  \begin{eqnarray}
    &&\lambda_{\vec q}=\frac{1+s\,q_1}{s_\perp q_2}\\
    &&q_1+s=\lambda_{\vec q}\,s\,s_\perp\,q_2\\
    &&q_2s=\lambda_{\vec q}s_\perp(s-q_1)\label{cond3_1}\\
    &&q_2=\lambda_{\vec q}s_\perp(1-sq_1).\label{cond4_1}
  \end{eqnarray}
\end{subequations}
This set of four equations for three variables $q_1$, $q_2$ and $\lambda_{\vec q}$ is non-contradictory
when two of these equations are linearly dependent. When the state is pure ($s=1$), 
the last two equations (\ref{cond3_1}) and (\ref{cond4_1}) are equivalent and the solution is $q_1^2+q_2^2=1$. Thus for
pure states there is a continuous set of optimal POVMs (\ref{povm}) 
parametrized by a vector $\vec q$ which lies on a circle of unit radius. 
If the state is mixed ($s<1$), the last two
equations are non-contradictory only if $q_1=0$ and the other two equations give $q_2=\pm1$. 
This is a dramatic difference when compared to the pure state case -- a continuous set of POVMs reduces to just two possibile projection operators, see Fig.~\ref{vecs_fig}. 

Finally, we point that when the logarithmic derivative is known, some subset of optimal measurements is given by the projection operators onto the eigen-states 
of $\hat{\mathcal{L}}_{\hat\rho}$. In the case of a single qubit, this procedure gives only the $q_1=0$, $q_2=\pm1$ projectors, even for pure states.

\subsection{Estimation from the population imbalance}

In an $N$-qubit MZI, usually the phase is estimated from the population imbalance between the two arms of the interferometer. Here we show under which conditions this estimation
strategy is optimal in the simplest one-qubit case. 

The population imbalance POVMs are two operators
\begin{equation}
  \hat E_+=|+\rangle\langle+|\ \ \ \mathrm{and}\ \ \ \hat E_-=|-\rangle\langle-|.
\end{equation}
which project the state $\hat\rho$ on either of the two modes.
The corresponding probabilities of detecting the qubit in one of the arms are
\begin{equation}
  p_{\pm}=\tr(\hat\rho\,\hat E_{\pm})=\frac12(1\pm s_z)
\end{equation}
According to Eq.~(\ref{cfi}) and using the $\theta$-dependence of the vector $\vec s$ from Eq.~(\ref{rotated}), we obtain that the CFI for the estimation from the population imbalance is
\begin{equation}\label{imb}
  F_{\rm imb}=\frac1{p_+}\left(\frac{\partial p_+}{\partial\theta}\right)^2+\frac1{p_-}\left(\frac{\partial p_-}{\partial\theta}\right)^2=\frac{s_x^2}{1-s_z^2}.
\end{equation}
This CFI saturates the bound of the QFI from Eq.~(\ref{qfi}) when either (a) $s_x^2+s_z^2=1$ or (b) $s_z=0$. When (a) is true, the state is pure, and the CFI does not depend on $\theta$.
In other words, for any initial pure state $\hat\rho_{\rm in}$ lying in the $x-z$ plane, the population imbalance measurement is optimal for the phase estimation. When the state is mixed,
only (b) can be true and then for every $\theta$ there is only one orientation of  $\hat\rho_{\rm in}$, which gives $s_z=0$. 

This once again shows how the optimal estimation strategies change abruptly as soon as the state is mixed. While there is a continuum of pure states, which when used for the
phase estimation from the population imbalance give maximal value of the CFI, there is only one such mixed state for each $\theta$. 

\section{Two qubits}
\label{wern}

In this section, we extend the analysis of the optimal estimation strategies to two spin-$\frac12$ bosons. A general density matrix of such system with its 
eight independent parameters, it too difficult to investigate. However there is a set of ``Werner'' states \cite{WernerPRA1989}, 
which are described with a single real coefficient $\alpha$. These states have a particularly simple form
\begin{equation}\label{wer}
  \hat\rho_{\rm w}=\frac{1-\alpha}3\hat{\mathds1}+\alpha\,\hat\Pi_{\rm TF},
\end{equation}
where $\hat\Pi_{\rm TF}=\big|1,1\big\rangle\big\langle1,1\big|$ is a projection over the Twin-Fock state. The ket $\ket{1,1}$ denotes a state where each mode of the interferometer is occupied by
one particle.

When $\alpha$ varies from $0$ to $1$, $\hat\rho$ changes from a complete mixture, which is useless for the parameter estimation, to a strongly entangled pure Twin-Fock state,
which provides sub SNL in the Mach-Zehnder interferometer \cite{HollandPRL1993}.
Werner states are a narrow subset of all possible two spin-$\frac12$ bosonic states, nevertheless -- as we show below -- 
they provide valuable insight into the optimal estimation strategies in quantum metrology. 

In analogy to the previous section, we use a generic linear interferometric transformation 
\begin{equation}\label{trans_2q}
  \hat U(\theta)=\exp\left[-i\theta\,\vec n\cdot\vec{\hat J}\right].
\end{equation}
The ``composite'' angular momentum operators are a sum of corresponding single-particle Pauli matrices, i.e. $\hat J_i=\frac12\hat\sigma_i^{(1)}+\frac12\hat\sigma_i^{(2)}$, where $i=x,y,z$ 
and the upper index labels the particles. 

First, we calculate the QFI, using the expression from Eq.~(\ref{fq_gen}). The Werner state written in the mode occupation basis is already diagonal and reads
\begin{eqnarray}
  \hat\rho_{\rm w}=\left(\begin{array}{ccc}
    \frac{1-\alpha}3&0&0\\
    0&\frac{1+2\alpha}3&0\\
    0&0&\frac{1-\alpha}3
    \end{array}\right).
\end{eqnarray}
Since the generator of the phase transformation from Eq.~(\ref{trans_2q}) is $\hat h=\vec n\cdot\vec{\hat J}$, then according to Eq.~(\ref{fq_gen}) evaluation of the QFI is straightforward and we obtain
\begin{equation}
  F_Q(\alpha)=\frac{12 \alpha^2}{2+\alpha}\left(n_x^2+n_y^2\right)
\end{equation}
Note that the $z$-component of the generator does not contribute to the QFI, because $\hat\rho_{\rm w}$ is invariant upon rotation around the $z$ axis. The QFI depends only on the length of the projection
of the vector $\vec n$ onto the $x$-$y$ plane.
Therefore, it is reasonable to consider only such transformations, which lie in the $x$-$y$ plane. 
In this case, the value of the QFI depends only on the length of the vector $\vec n$. Thus, without any loss of generality in the remaining of this Section we can restrict
to the MZI transformation $\hat U(\theta)=\exp\left[-i\theta\hat J_y\right]$. If so, the QFI is simply equal to
\begin{equation}\label{fq_2}
  F_Q(\alpha)=\frac{12 \alpha^2}{2+\alpha}. 
\end{equation}
As anticipated at the beginning of this section, $F_Q(0)=0$. If the Werner state is transformed by a collective rotation as in Eq.~(\ref{trans_2q}), 
then according to the criterion of the QFI for this 
interferometer, the symmetric Werner state is usefully entangled, when $\alpha\geqslant\frac23$. In the extreme case, for a Twin-Fock state $\alpha=1$, we
obtain the Heisenberg scaling, i.e. $F_Q(1)=4$. This result can be compared with the concurrence \cite{horo}, 
the entanglement measure for two qubits, which tells that the state (\ref{wer}) is entangled already when
$\alpha>\frac14$. This example confirms a known fact that not all entangled states are usefully entangled for interferometric transformations as in Eq.~(\ref{trans_2q}). 

\subsection{Optimal measurements}

In the next step, we find the optimal measurements, which saturate the bound set by the QFI (\ref{fq_2}). To this end, we determine the logarithmic derivative using Eq.~(\ref{defL}). 
A simple calculation gives that
\begin{equation}
  \hat{\mathcal L}_{\hat\rho}=-\frac{6i}{2+\alpha}\left[\hat J_y,\hat\rho_{\rm w}(\theta)\right],
\end{equation}
where $\hat\rho_{\rm w}(\theta)=e^{-i\theta\hat J_y}\hat\rho_{\rm w}e^{i\theta\hat J_y}$.
In analogy to the single-qubit case, we should now parametrize the POVMs similarly as in Eq.~(\ref{povm}) and find the parameters from the condition Eq.~(\ref{opt_cond}). 
However, this procedure gives equations, which cannot be solved in a simple way. Therefore, we restrict to those optimal POVMs, which can be found by the diagonalization of the logarithmic
derivative. We write down $\hat{\mathcal L}_{\hat\rho}$ in the matrix form
\begin{eqnarray*}
  \hat{\mathcal L}_{\hat\rho}=\frac{6\alpha}{\sqrt2(2+\alpha)}\left(\begin{array}{ccc}
      \frac1{\sqrt2}\sin2\theta&-\cos2\theta&-\frac1{\sqrt2}\sin2\theta\\
      -\cos2\theta&-\sqrt2\sin2\theta&\cos2\theta\\
      -\frac1{\sqrt2}\sin2\theta&\cos2\theta&\frac1{\sqrt2}\sin2\theta
    \end{array}\right)
\end{eqnarray*}
and obtain a set of three eigen-states
\begin{subequations}\label{2qbt}
  \begin{eqnarray}
    &&\!\!\!\!\!\!\!\!\!|\Psi_1\rangle=\frac{\left(\cos\theta-\sin\theta\right)}{\sqrt2}|\psi_-\rangle-\frac{\left(\cos\theta+\sin\theta\right)}{\sqrt2}|1,1\rangle\\
    &&\!\!\!\!\!\!\!\!\!|\Psi_2\rangle=\frac{\left(\cos\theta+\sin\theta\right)}{\sqrt 2}|\psi_-\rangle+\frac{\left(\cos\theta-\sin\theta\right)}{\sqrt2}|1,1\rangle\\
    &&\!\!\!\!\!\!\!\!\!|\Psi_3\rangle=|\psi_+\rangle,
  \end{eqnarray}
\end{subequations}
where $|\psi_{\pm}\rangle=\frac{|2,0\rangle\pm|0,2\rangle}{\sqrt 2}$.
The optimal measurements depend on $\theta$, have a complicated form. Notice that when $\theta=0$ the following transformation
\begin{equation}\label{v}
  \hat V=\exp\left(i\frac{\pi}{2}\frac{\hat J_x \hat J_y + \hat J_y \hat J_x}{2}\right)\exp\left( i \frac{\pi}{4} \hat J_y\right).
\end{equation}
is applied to Eq. (\ref{2qbt}), it results in 
$\hat V |\Psi_1\rangle = |0,2\rangle$, $\hat V |\Psi_2\rangle = |1,1\rangle$ and $\hat V |\Psi_3\rangle = |2,0\rangle$. In this way, we obtain the eigenstates of the $\hat J_z$ operator,
and the optimal measurement is based on the determination of the population imbalance. Nevertheless, to accomplish this we needed an additional operation (\ref{v}) on the state. 
This transformation is non-local -- it correlates the particles, since the product of two angular-momentum operators cannot be written as a sum of operators acting 
on each qubit independently.

\subsection{Estimation from the population imbalance}

\begin{figure}[htb!]
  \includegraphics[clip, scale=0.35]{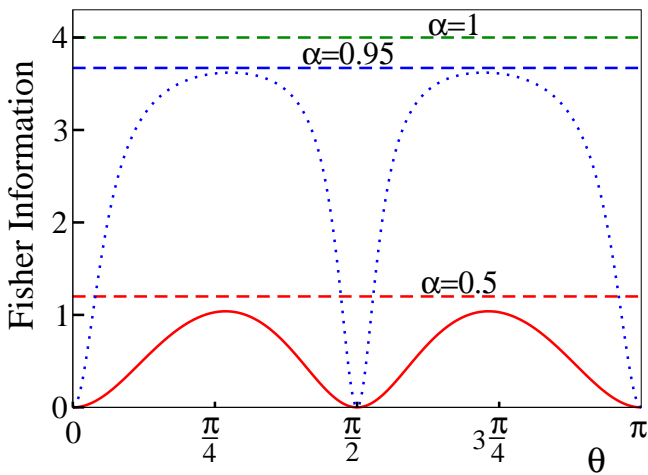}
  \caption{(Color online) The CFI given by Eq.~(\ref{imb_res}) for $\alpha=0.5$ (solid red), $\alpha=0.95$ (dotted blue) and $\alpha=1$ (dashed green) as a function of $\theta$. 
    The dashed lines denote the corresponding values of the QFI (\ref{fq_2}), which for $\alpha=1$ is equal to the CFI.
  }\label{werner_fig}
\end{figure}

We now consider a common estimation strategy based on the measurement of the imbalance of the population of the two modes. The CFI defined in Eq.~(\ref{cfi}) is a sum of three terms,
\begin{equation}\label{cfi_2}
  F_{\rm imb}=\sum_{n=0}^2\frac1{p(n|\theta)}\left(\frac{\partial p(n|\theta)}{\partial\theta}\right)^2,
\end{equation}
where $p(n|\theta)$ is a probability for finding $n$ particles in one of the modes and $2-n$ in the other and is given by
\begin{equation}
  p(n|\theta)=\tr\Big[|n,N-n\rangle\langle n,N-n|\hat\rho_{\rm w}(\theta)\Big].
\end{equation}
Using the the density matrix of the Werner states from Eq.~(\ref{wer}) we obtain that $p(0|\theta)=p(2|\theta)=\frac{1-\alpha}3+\frac\alpha2\sin^2\theta$
and $p(1|\theta)=\frac{1-\alpha}3+\alpha\cos^2\theta$, which put into (\ref{cfi_2}) gives
\begin{equation}\label{imb_res}
  F_{\rm imb}=\frac{36\,\alpha^2\,\sin^2(2\theta)}{[4-\alpha(1+3\cos(2\theta))][2+\alpha(1+3\cos(2\theta))]}.
\end{equation}
Only for $\alpha=1$, when the Werner state is pure, 
 $F_{\rm imb}=4$, so it does not depend on $\theta$ and saturates the bound of the QFI. As shown in Fig.~\ref{werner_fig}, for other values
of $\alpha$ the estimation from the population imbalance is non-optimal for all values of $\theta$.

This is another example, after the single qubit case, of how the estimation strategy, which is optimal for a pure state, immediately deteriorates as soon as the state becomes mixed.

\section{Optimal measurements for N-qubit pure states}\
\label{Npure}

So far, we have identified the optimal measurements for one and two qubits. It is natural to generalize these results and ask which are the optimal 
measurements for $N$ qubits undergoing a linear interferometric transformation. However, the methods used in the previous two sections, which employed the 
logarithmic derivative $\hat{\mathcal L}_{\hat\rho}$, cannot be extended to higher $N$. This is because the mere analytical determination of $\hat{\mathcal L}_{\hat\rho}$ becomes 
cumbersome.
Nevertheless, as we show below, for pure $N$-qubit states, some optimal estimation strategies can be found using the notion of the statistical distance. 

\subsection{QFI and the statistical distance}

In \cite{braun} it is shown how the QFI from Eq.~(\ref{fq_gen}) is related to the statistical distance between two neighboring states \cite{wootters}. 
Although this general result is valid whenever
a parameter is estimated from measurements performed on a $\theta$-dependent state $\ket\psi$, below we present this formalism in context of quantum interferometry. 

\begin{figure}[htb!]
  \includegraphics[clip, scale=0.4]{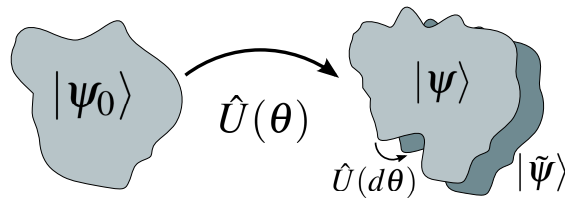}
  \caption{(Color online)
    Schematic representation of the two steps performed to calculate the QFI. The input state $\ket{\psi_0}$, which enters the interferometer, is transformed by a unitary evolution
    operator $\hat U(\theta)$, giving the output state $\ket\psi$.
    To calculate the speed at which the state changes, and thus the statistical distance, we make a further infinitesimal rotation $\hat U(\theta)$ to obtain $\ket{\tilde\psi}$. 
  }\label{fig_scheme}
\end{figure}

To this end, we follow the scheme pictured in Fig.~\ref{fig_scheme}. 
First, consider a pure two-mode {\it input} state $\ket{\psi_0}$, which is transformed by a unitary evolution operator $\hat U(\theta)=e^{-i\theta\hat J_k}$,
where $k=x,y,z$. As a result, we obtain a $\theta$-dependent {\it output} state, which when expanded in the basis of mode occupations reads
\begin{equation}
  \ket{\psi}=e^{-i\theta\hat J_k}\ket{\psi_0}=\sum_{j=0}^N \sqrt{p_j} e^{i \varphi_j} \ket{j}\equiv\sum_{j=0}^N C_j \ket j.\label{init}
\end{equation}
Here $\ket j$ is a Fock state with $j$ particles in the left and $N-j$ in the right arm. 
The neighboring state is found by applying a further infinitesimal transformation $e^{-id\theta\hat J_k}$, which gives
\begin{subequations}\label{act}
  \begin{eqnarray}
    \ket{\tilde\psi}&=&e^{-id\theta\hat J_k}\ket\psi\simeq\sum_{j=0}^N(1-id\theta\hat J_k)C_j\ket j\\
    &=&\sum_{j=0}^N(1-id\theta\eta^{(k)}_j)C_j\ket j\equiv\sum_{j=0}^N\tilde C_j\ket j,\label{act2}
  \end{eqnarray}
\end{subequations}
The coefficient $\eta^{(k)}_j$ is a result of acting with $\hat J_k$ on a ket $\ket j$ and is equal to
\begin{subequations}\label{eta}
  \begin{eqnarray}
    \eta^{(x)}_j&=&\frac12\,\frac{\alpha_jC_{j+1}+\alpha_{j-1}C_{j-1}}{C_j}\\
    \eta^{(y)}_j&=&\frac1{2i}\,\frac{\alpha_jC_{j+1}-\alpha_{j-1}C_{j-1}}{C_j}\\
    \eta^{(z)}_j&=&j-\frac N2,\label{etaz}
  \end{eqnarray}
\end{subequations}
where $\alpha_j=\sqrt{(j+1)(N-j)}$. The state (\ref{act2}) can be alternatively written as
\begin{equation}
  \ket{\tilde\psi}=\sum_{j=0}^N \sqrt{p_j+dp_j} e^{i (\varphi_j+d\varphi_j)} \ket{j}.\label{tilde}
\end{equation}
where the probability- and phase-increments are
\begin{subequations}
  \begin{eqnarray}
    &&dp_j = |\tilde C_j|^2 - |C_j|^2=2\,\im\,\eta^{(k)}_j|C_j|^2d\theta\\
    &&e^{id\varphi_j} = \frac{\tilde C_j}{|\tilde C_j|} \frac{|C_j|}{C_j}=e^{-i\,\re\,\eta^{(k)}_jd\theta}.
  \end{eqnarray}
\end{subequations}
The distance between two neighboring states is equal to
\begin{subequations}\label{PS2}
  \begin{eqnarray}
    &&ds_{\mathrm{ps}}^2=1-|\braket\psi{\tilde\psi}|^2\\
    &&=\sum_{j=0}^N \frac{dp_j^2}{p_j}  + 4\left[ \sum_{j=0}^N p_j d\varphi_j^2 -  \left( \sum_{j=0}^N p_j d\varphi_j \right)^2 \right]\\
    &&\equiv\sum_{j=0}^N \frac{dp_j^2}{p_j}+4\Delta^2 d\varphi.
  \end{eqnarray}
\end{subequations}
Finally, the QFI can be interpreted as the speed at which the state changes upon the infinitesimal increment of the parameter $\theta$ and therefore it reads 
\begin{subequations}\label{gen}
  \begin{eqnarray}
    \label{gen1}\!\!\!\!&&F_Q=\frac{ds_{\mathrm {ps}}^2}{d\theta^2}=4\sum_{j=0}^N|C_j|^2\left(\im\,\eta^{(k)}_j\right)^2+\\
    \label{gen2}&&4\sum_{j=0}^N|C_j|^2\left(\re\,\eta^{(k)}_j\right)^2-4\left(\sum_{j=0}^N|C_j|^2\re\,\eta^{(k)}_j\right)^2.
  \end{eqnarray} 
\end{subequations}
The above result is equivalent to the well known expression for the QFI for pure states $F_Q=4\Delta^2\hat J_k$. 
However, as will become evident below, for the purpose of finding the optimal measurements, 
it is more convenient to keep the QFI in the form of Equations (\ref{gen}), i.e. as a sum of two non-negative parts --  
the change of the probability $p_j$ and the variance of the phase increment $d\varphi_j$. 
Namely, if the latter term is zero, then the CFI calculated 
using the probability $p_j=|\braket{j}{\psi}|^2$ of finding the system in state $\ket{j}$ is equal to QFI. This means that no information about $\theta$ is carried by the phases $\varphi_j$, which 
 are not witnessed by the projection measurement $\ket{j}\bra{j}$ and thus do not contribute to probabilities $p_j$.

\subsection{``In-situ'' measurements -- localized modes}

In this section, by referring to Eq.~(\ref{gen}), we identify two optimal measurements performed ``in-situ'', 
when the particles remain trapped in the two arms of the interferometer and their spatial mode functions do not overlap. 

\subsubsection{Estimation from the full correlation}

As a first example, we consider the phase estimation from the full $N$-body probability
\begin{eqnarray}\label{pn}
  p_N(\x|\theta)&=&\frac1{N!}\bra\psi\hat\Psi^\dagger(x_1)\ldots\hat\Psi^\dagger(x_N)\hat\Psi(x_N)\ldots\hat\Psi(x_1)\ket\psi\nonumber\\
  &\equiv&\bra\psi\hat G(\x)\ket\psi.  
\end{eqnarray}
of finding particles at positions $\x=(x_1\ldots x_N)$. 
The two-mode
field operator is $\hat\Psi(x)=\psi_a(x)\hat a+\psi_b(x)\hat b$ and the wave-packets are separated in two arms of the interferometer, for instance by imposing
$\psi_{a}(x)=0$ for $x<0$ and $\psi_b(x)=0$ for $x>0$. The $\theta$-dependence of the probability $p_N(\x|\theta)$ comes from the state $\ket\psi$ from Eq.~(\ref{init}), which
is used to calculate the average value of the operator $\hat G(\x)$. 

The estimation sequence relies upon detecting positions of $N$ atoms in $m\gg1$ experiments. If the phase is then obtained from the maximum likelihood estimator, then according
to the Fisher theorem, its sensitivity is given by
\begin{equation}
  \Delta^2\theta=\frac1m\frac1{F_N},
\end{equation}
where $F_N$ is the CFI which is equal to
\begin{equation}\label{N_cfi}
  F_N=\int\!\! d\x\frac1{p_N(\x|\theta)}\left(\frac{\partial p_N(\x|\theta)}{\partial\theta}\right)^2.
\end{equation}
In order to calculate $F_N$ we first evaluate the derivative of the probability (\ref{pn}),
\begin{eqnarray}\label{der}
  &&\partial_\theta p_N(\x|\theta)=i\bra\psi\hat J_k\hat G(\x)\ket\psi-i\bra\psi\hat G(\x)\hat J_k\ket\psi=\\
  &&2\,\im\,\bra\psi\hat G(\x)\hat J_k\ket\psi=2\,\im\!\!\!\sum_{j,j'=0}^N\!\!\!C_j^*C_{j'}\eta_{j'}^{(k)}\bra j\hat G(\x)\ket{j'}\nonumber.
\end{eqnarray}
The CFI is therefore equal to
\begin{equation}
  F_N=4\int\!\! d\x\ 
  \frac{\left[\im\!\!\!\sum\limits_{j,j'=0}^N\!\!\!C_j^*C_{j'}\eta_{j'}^{(k)}\bra j\hat G(\x)\ket{j'}\right]^2}{\im\!\!\!\sum\limits_{j,j'=0}^N\!\!\!C_j^*C_{j'}\bra j\hat G(\x)\ket{j'}}.
\end{equation}
We now define $\Omega_\mu$ by saying that $\x\in\Omega_\mu$ when $x_1\ldots x_\mu<0$ and $x_{\mu+1}\ldots x_N>0$. Using this definition we obtain
\begin{equation*}
  F_N\!=\!4\sum_{\mu=0}^N\!\binom N\mu\int\limits_{\x\in\Omega_\mu}\!\!\!\! d\x\ \!
  \frac{\left[\im\!\!\!\sum\limits_{j,j'=0}^N\!\!\!C_j^*C_{j'}\eta_{j'}^{(k)}\bra j\hat G(\x)\ket {j'}\right]^2\!\!}{\im\!\!\!\sum\limits_{j,j'=0}^N\!\!\!C_j^*C_{j'}\bra j\hat G(\x)\ket {j'}}
\end{equation*}
where the combinatory factor is due to indistinguishability of particles and stands for all possible choices of $\mu$ particles out of a set of $N$. When $\x\in\Omega_\mu$,
then for separated wave-packets $\hat G(\x)\ket n\propto\frac{\mu!(N-\mu)!}{N!}\ket n\delta_{n\mu}$ and the above integral gives
\begin{equation}\label{cfiN_res}
  F_N=4\sum_{j=0}^N|C_j|^2\left(\im\,\eta^{(k)}_j\right)^2.
\end{equation}
We notice that this expression is equal to the first line of the QFI, see (\ref{gen1}). 

Therefore, estimation from the $N$-body probability of trapped particles is optimal only if the other terms in line (\ref{gen2}) vanish, which requires $\re\,\eta^{(k)}_j\equiv0$ for all 
$j$. According to Equations (\ref{eta}), this condition can be satisfied only for the rotations around $x$ and $y$ axes. In the other case,  
$\im\,\eta^{(z)}_j=0$ and thus $F_N=0$, because the simple phase imprint $e^{-i\theta\hat J_z}$ requires further mode mixing to provide information about $\theta$. 

For the rotation around $x$-axis, $\re\,\eta^{(x)}_j=0$ if $C_j=i^ja_j$, while for the $y$-axis the condition is
$C_j=e^{i\phi}a_j$, where $a_j\in\mathbb{R}$ and $\phi$ is a common phase. In particular for $\phi=0$, the measurement is optimal, when all $C_j$'s are real. 
Since the elements of the Wigner rotation matrix -- which transforms the input state $\ket{\psi_0}$ into the output state $\ket\psi$ -- are all real  \cite{nota_wig}, 
we conclude that if the input state of the MZI has real coefficients, the estimation from $p_N$ is optimal.

\subsubsection{Estimation from the population imbalance}
\label{imb_pure}

Although phase estimation from the $N$-body probability is optimal for a wide class of states and rotations around $x$ and $y$,
it has one major flaw -- it is unpractical, since it requires sampling of a vast configurational space. We now show, that the same value of the CFI as in Eq.~(\ref{cfiN_res})
can be obtained, when the phase is estimated from a simple population imbalance measurement. 

The probability of having $j$ atoms in the mode $a$ and $N-j$ in $b$ is
\begin{equation}\label{p_imb}
  p(j|\theta)=|\braket j\psi|^2=|C_j|^2.
\end{equation}
Similarly as in Eq.~(\ref{der}), its derivative reads
\begin{equation}
  \partial_\theta p(j|\theta)=2|C_j|^2\im\,\eta_j^{(k)}.
\end{equation}
Therefore, the CFI calculated with (\ref{p_imb})
\begin{equation}
  F_{\rm imb}=\sum_{j=0}^N\frac1{p(j|\theta)}\left(\frac{\partial\, p(j|\theta)}{\partial\theta}\right)^2=4\sum_{j=0}^N|C_j|^2\left(\im\,\eta^{(k)}_j\right)^2
\end{equation}
is equal to (\ref{cfiN_res}). In consequence, the QFI from Eq.~(\ref{gen}) is saturated with the same family of states for the $x$ and $y$ rotations as in the case of
the estimation from $p_N(\x|\theta)$. Note that this result, obtained independently in \cite{caves_new}, is a step forward with respect
to the work of Hofmann \cite{hof}, where the saturation of the QFI bound with the population imbalance measurement was reported for the MZI and symmetric states with  $C_{j}=C_{N-j}$.

\subsection{Measurement after expansion}

As argued above, when the interferometer rotates the state around the $z$-axis, giving a sole phase-imprint, further manipulation is necessary to exchange the
information about the phase between the two modes. Here we assume, that this operation is realized by letting the two mode functions $\psi_a(x)$ and $\psi_b(x)$
expand and form an interference pattern. In such situation, the two modes cannot be distinguished anymore, and it is not possible to define a proper population imbalance operator.
Instead, one must estimate $\theta$ in some different way. For instance, estimation from the least-squares fit of the one-body probability 
$p_1(x|\theta)=\frac1N\bra\psi\hat\Psi^\dagger(x)\hat\Psi(x)\ket\psi$ to the interference pattern, 
although gives sub shot-noise sensitivity when the input state $\ket{\psi_0}$ is phase-squeezed \cite{grond}, is never optimal \cite{chwed_njp2}. 

Nevertheless, the optimal measurement can be identified and it is the $N$-body CFI from Eq.~(\ref{N_cfi}) which saturates the bound of the QFI under following additional assumptions
\cite{chwed_njp1}. First, the information between the two modes must be fully exchanged. This means, that the envelopes of $\psi_a(x)$ and $\psi_b(x)$ fully overlap and the functions
only differ by the phase. This is true if initially $\psi_a(x)$ and $\psi_b(x)$ are of the same shape but are separated in space and then expand to reach the far-field regime.
Another requirement is that the coefficients of the initial state
$C_j^{(0)}$
are real and posses the symmetry $C_j^{(0)}=C_{N-j}^{(0)}$. States having these properties naturally appear in the context of quantum interferometry with 
ultra-cold gas trapped in the double-well potential. Namely, the ground state of the symmetric two-mode Bose-Hubbard Hamiltonian for every ratio of the interaction strength
$U$ to the tunneling rate $J$ has real and symmetric coefficients $C_j^{(0)}$.

According to Eq.~(\ref{init}), the rotation around the $z$ axis transforms the state into
\begin{equation}
  \ket\psi=e^{-i\theta\hat J_z}\ket{\psi_0}=\sum_{j=0}^NC_j\ket j,
\end{equation}
where $C_j=C_j^{(0)}e^{-i\theta\left(j-\frac N2\right)}$. As argued in detial in \cite{chwed_njp1}, the CFI from Eq.~(\ref{N_cfi}) can be calculated under the aforementioned assumptions
and the outcome is
\begin{equation}
  F = 4 \sum_j |C_j|^2 \eta_j^{(z)2} = 4 \Delta^2\hat{J}_z = F_Q,
\end{equation}
where $\eta_j^{(z)}$ was defined in Eq.~(\ref{etaz}). This shows that the estimation from the $N$-th body correlation in the far field is optimal.

\section{Conclusions}
\label{concl}

In this work we have identified optimal measurements in various two-mode interferometric system. For a single
qubit, we have shown that a continuous set of optimal estimation strategies for pure state reduces to only two possibilities when $\hat\rho_0$ is mixed. This simple example shows, that
the problem of finding optimal measurements is very sensitive to the variations in the input state $\hat\rho_0$. We then switched to two qubits and considered an entangled
two-mode symmetric Werner state, which depending on parameter $\alpha$ can be either pure $\alpha=1$ or mixed $\alpha<1$. Again we have derived an expression for the optimal 
measurement, which requires projecting the output state onto the basis of entangled states of two particles. We have also shown that estimation from the measurement of the population 
imbalance between the two arms of the interferometer is optimal only for a pure state $\alpha=1$.
Finally, we considered pure states of $N$ qubits and shown which states $\hat\rho_0$ allow to reach the bound of the QFI using the measurement of the population
imbalance or the $N$-th order correlation function. 

\section{Acknowledgements}
J. Ch. acknowledges the Foundation for Polish Science International TEAM Programme co-financed by the EU European Regional Development Fund and the support of 
the Polish NCBiR under the ERA-NET CHIST-ERA project QUASAR.
T.W. acknowledges the Foundation for Polish Science International Ph.D. Projects Programme co-financed by the EU European Regional Development Fund and
the National Science Center grant no. DEC-2011/03/D/ST2/00200. 
L.P. acknowledges financial support by MIUR through FIRB Project No. RBFR08H058.
QSTAR is the MPQ, IIT, LENS, UniFi joint center for Quantum Science
and Technology in Arcetri.
This research was partially supported by the EU-STREP Project QIBEC.

\end{document}